\documentclass[a4paper,aps,preprintnumbers,showpacs,twocolumn,superscriptaddress,nofootinbib,amsmath,amssymb]{revtex4-1}
\usepackage[dvips]{graphics}

\def\a{\widetilde{\alpha}}

\newcommand{\rem}[1]{}
\begin{document}
\title{Are eikonal quasinormal modes linked to the unstable circular null geodesics?}

\author{R. A. Konoplya}
\affiliation{Theoretical Astrophysics, IAAT, Eberhard-Karls University of T\"ubingen, T\"ubingen 72076, Germany}
\affiliation{Institute of Physics and Research Centre of Theoretical Physics and Astrophysics,
Faculty of Philosophy and Science, Silesian University in Opava, CZ-746 01 Opava, Czech Republic}

\author{Z. Stuchl\'{i}k}
\affiliation{Institute of Physics and Research Centre of Theoretical Physics and Astrophysics,
Faculty of Philosophy and Science, Silesian University in Opava, CZ-746 01 Opava, Czech Republic}

\begin{abstract}
In Phys. Rev. D 79, 064016 (2009) it was claimed that quasinormal modes which any stationary, spherically symmetric and asymptotically flat black hole emits in the eikonal regime are determined by the parameters of the circular null geodesic: the real and imaginary parts of the quasinormal mode are multiples of the frequency and instability timescale of the circular null geodesics respectively. We shall consider asymptotically flat black hole in the Einstein-Lovelock theory, find analytical expressions for gravitational quasinormal modes in the eikonal regime and analyze the null geodesics. Comparison of the both phenomena shows that the expected link between the null geodesics and quasinormal modes is violated in the Einstein-Lovelock theory. Nevertheless, the correspondence exists for a number of other cases and here we formulate its actual limits.
\end{abstract}

\maketitle

\section{Introduction}

Recent progress in black-hole observations in the gravitational and electromagnetic spectra as well as theoretical efforts to test strong gravity via black holes \cite{Abbott:2016blz,TheLIGOScientific:2016src,Goddi:2016jrs,Bambi:2015kza,Konoplya:2016pmh} makes it important to understand possible correlations between characteristics of both fields in the vicinity of a black hole. In \cite{Cardoso:2008bp} it was stated that parameters of the unstable circular null geodesics around any stationary  spherically symmetric and asymptotically flat black holes, such as  the angular velocity $\Omega_c$  and the principal Lyapunov exponent $\lambda$, are in the remarkable correspondence with the quasinormal modes \cite{QNMreviews} that the  black hole emits in the eikonal (short wavelengths or high multipole number $\ell$) part of its spectrum. There it was shown that the eikonal quasinormal frequencies of the four and higher dimensional Schwarzschild black hole are
\begin{equation}\label{QNM}
\omega_n=\Omega_c\,\ell-i(n+1/2)\,|\lambda|,
\end{equation}
where $n$ is the overtone number. In addition, it was argued that the above formula must be valid for all stationary, spherically symmetric non-asymptotically flat black holes, allowing for the outgoing wave boundary condition in the far region (for example, asymptotically de Sitter black holes).

The issue of rotating black holes was also addressed in  \cite{Cardoso:2008bp}. For slowly rotating black holes, the eikonal real oscillation   frequencies are  linear  combinations  of  the  orbit’s  precessional  and  orbital  frequencies, while for Kerr black holes of arbitrary spin the link between photon spheres and eikonal quaisnormal modes is more complicated  \cite{Yang:2012he}.  At the same time it has been recently noticed that the association of the characteristics of the null geodesics with quasinormal modes is more based on the history of the specific black-hole models than an actual and generic constraining link \cite{Khanna:2016yow}. The essential element of the correspondence is the event horizon: when the event horizon is replaced by the reflecting surface \cite{Price:2017cjr} or a wormhole throat \cite{Khanna:2016yow}, the correspondence (\ref{QNM}) is not observed.

The arguments of \cite{Cardoso:2008bp} for spherically symmetric black holes implied the applicability of the WKB formula developed in \cite{Schutz:1985zz} for a particular, though quite wide, class of effective potentials, which have the form of the potential barrier with a single extremum outside the event horizon and approach constant values at the horizon and spacial infinity (or de Sitter horizon). This requirement certainly cannot be guaranteed ad hoc, so that, if one supposes that this initial setting is not valid for some black hole, then the counterexample would be straightforward. At the same time, there are a number of cases where the correspondence do works and even more cases where it is erroneously believed to be working (examples of both can be found in \cite{EikonalWork,Hod,Gallo} and references therein). Therefore, here we are interested in testing the possible correspondence in even the narrower setup: Assuming that radiation of gravitational waves by a spherical black hole is governed by ``the WKB-well-behaved'' effective potential with a single extremum, we would like to learn how broad the set of situations is, in which the relation (\ref{QNM}) between null geodesics and quasinormal modes is guaranteed?

With this aim we shall consider the situation when the WKB formula is accurate and even \emph{exact} in the eikonal regime, and, nevertheless, the relation (\ref{QNM}) is not fulfilled. We shall show that there is a counterexample (suggested by the Einstein-Lovelock theory) to the claimed correspondence. The Einstein-Lovelock theory of gravity \cite{Lovelock:1971yv} is the most general mathematically consistent metric theory, leading to second order equations of motion in arbitrary number of spacetime dimensions $D$. It is natural generalization of Einstein theory in  $D >4$ and may represent string theory motivated quantum corrections to the classical geometry in higher dimensions. Thus, this discussion gives us also an excuse to find  analytic formulas for the eikonal quasinormal modes for gravitational perturbations of higher curvature corrected black holes and complement, in this way, a recent WKB analysis of quasinormal spectrum of Lovelock black holes, which was  done in \cite{Yoshida:2015vua}.

The paper is organized as follows: Sec. II gives the basic formulas for calculations of the principal Lyapunov exponent and the angular velocity for the unstable null geodesics in spherically symmetric spacetimes. In Sec. III the Lyapunov exponents and the angular velocity are found for the asymptotically flat Einstein-Gauss-Bonnet black hole. Sec. IV shows that the frequencies predicated by the Lyapunov exponent and angular velocity are different from those given by the WKB formula for the generic Einstein-Lovelock black hole. In Sec. V,  analytical formulas for quasinormal modes in the eikonal (i.e. high multipole numbers $\ell$) regime are written down in terms of black-hole parameters for the Einstein-Gauss-Bonnet case.
In Sec. VI we discuss the obtained results and formulate actual limits of the correspondence.

\section{Null geodesics in the background of spherically symmetric black holes}

A static, spherically symmetric metric in $D$ -dimensional spacetime has the form:
\begin{equation}\label{metric}
d s^2 = f(r) d t^2 - \frac{1}{g(r)} d r^2 -r^2 d\Omega_{n}^2,
\end{equation}
where the functions $f(r)$ and $g(r)$ represent solutions of the field equations under consideration and $d\Omega_n^2$ is a $(n=D-2)$-dimensional sphere.  Let us consider geodesic particle motion around such a black hole and restrict attention to stability of null circular orbits. The stability can be analyzed in terms of the so called Lyapunov exponents \cite{Lyapunov}. This kind of analysis for the Schwarzschild black hole was developed for the first time in \cite{Cornish:2001jy}.  In \cite{Dettmann} it was shown that when  a system consisting of any finite number of particles moves under the action of a scalar potential at a constant kinetic energy, then the Lyapunov exponents come in pairs which sum to the same constant. The equations of motions can be written in the following schematic way
\begin{equation}
\frac{d X_{i}}{d t} = H_{i}(X_{j}).
\end{equation}
A small deviation from a given orbit to a nearby curve through the small perturbation $\delta X_{i}$,
\begin{equation}
X_{i} \rightarrow X_{i} + \delta X_{i},
\end{equation}
implies the linearization of the equation of motion
\begin{equation}
\frac{d \delta X_{i}(t)}{dt} = K_{ij}(t) \delta X_{j}(t),
\end{equation}
where
\begin{equation}
K_{ij}(t) = \left.\frac{\partial H_{i}}{\partial X_{j}} \right\vert_{X_{i}(t)}
\end{equation}
 is called the \emph{infinitesimal evolution matrix}. The solution to the linearized equation can be expressed in terms of the \emph{evolution matrix} $L_{ij}$:
\begin{equation}
\delta X_{i} (t) = L_{ij} (t) \delta X_{j}(0).
\end{equation}
The evolution matrix obeys the relations
\begin{equation}
\dot{L}_{ij} (t) = K_{im} L_{mj} (t), \quad L_{ij} (0) = \delta_{ij}.
\end{equation}

The principal Lyapunov exponents are given by
\begin{equation}
\lambda = \lim_{t \rightarrow \infty} \frac{1}{t} \left(\frac{L_{jj}(t)}{L_{jj}(0)}\right).
\end{equation}
The general conditions for the existence of the above limit are given by the Oseledets theorem \cite{Oseledets}.
Following Cardoso et. al. \cite{Cardoso:2008bp}, one can see that the principal Lyapunov exponent for null geodesics around a static,  spherically symmetric metric (\ref{metric}) is
\begin{equation}\label{GenLyap}
\lambda = \frac{1}{\sqrt{2}}\sqrt{-\frac{r_c^2}{f_c}\left(\frac{d^2}{dr_*^2}\frac{f}{r^2}\right)_{r=r_c}},
\end{equation}
where the tortoise coordinate is defined as
$dr/dr_*=\sqrt{g(r)f(r)}$.
The coordinate angular velocity for the null geodesics is
\begin{equation}\label{angularvel}
\Omega_c = \frac{f_c^{1/2}}{r_c},
\end{equation}
where $r_{c}$ is the radius of the circular null geodesics,  satisfying the equation
\begin{equation}\label{circulareq}
 2f_c=r_cf'_c.
\end{equation}
With the above formulas at hand, one is able to analyze stability and angular velocity of particles orbiting around arbitrary static spherically symmetric black hole. Recent discussion of the general features and instabilities of the null geodesics in the arbitrary spherically symmetric spacetimes and Lyapunov exponents has been suggested in \cite{Jia:2017nen}.

\section{Null geodesics in the background of the Einstein-Gauss-Bonnet black hole}
Here we shall consider the null geodesics in the black-hole background within the Einstein-Gauss-Bonnet theory. The Lagrangian of the D-dimensional Einstein-Gauss-Bonnet theory has the form:
\begin{equation}\label{gbg3}
  \mathcal{L}=-2\Lambda+R+ k (R_{\mu\nu\lambda\sigma}R^{\mu\nu\lambda\sigma}-4\,R_{\mu\nu}R^{\mu\nu}+R^2),
\end{equation}
where $k= \alpha/((D-3)(D-4))$.
The metric function of the asymptotically flat Einstein-Gauss-Bonnet black hole is given by \cite{Boulware:1985wk}
\begin{equation}\label{metricfunction}
f(r) = g(r) = 1+ \frac{r^2}{2 \alpha} - \frac{r^2}{2 \alpha} \sqrt{1 + \frac{8 \alpha \mu}{(D-2) r^{D-1}}},
\end{equation}
where $\mu$ is the mass parameter.
It is well known that Gauss-Bonnet black holes, as well as their Lovelock generalizations, are gravitationally unstable when the coupling constant $\alpha$ (and higher order constants in the case of the Lovelock theory) are not small enough.
Therefore, in order to obtain concise and easily interpretable analytical expressions, one can expand all the necessary relations in terms of small parameter $\alpha$. Thus, the radius of the circular geodesics can be written as follows
\begin{equation}\label{rexpansion1}
r_{c} = r_{c0} + r_{c1} \alpha + r_{c2} \alpha^2 + {\cal O}(\alpha^3).
\end{equation}
When expanding in terms of the Gauss-Bonnet coupling, from here and on we shall imply that the corresponding dimensionless parameter is $\alpha/r_{H}^2$, where $r_{H}$ is the black hole radius. In order to measure everything in terms of the black-hole radius it is sufficient to re-parameterize the mass $\mu$ as a function of radius $r_{H}$ as in eq. 4 of \cite{Konoplya:2010vz}.
The equation for the null circular orbits (\ref{circulareq}) for the metric (\ref{metricfunction}) reads
\begin{equation}\label{circulareq2}
r^3 \mu(1 - D)+r^{D/2} \sqrt{(D-2) \left((D-2) r^D+8 r \alpha  \mu \right)}=0.
\end{equation}
Substituting  (\ref{rexpansion1}) into (\ref{circulareq2}), one can find the coefficients of the expansion  (\ref{rexpansion1}), which are:
$$ r_{c0} = \left(\frac{D-2}{(D-1) \mu }\right)^{\frac{1}{3-D}}, \quad r_{c1} = -\frac{4 \left(\frac{D-2}{(D-1) \mu  }\right)^{\frac{1}{D-3}}}{D^2-4 D+3}, $$
\begin{equation}\label{rc0}
r_{c2} = -\frac{24 \left(\frac{D-2}{(D-1) \mu }\right)^{\frac{3}{D-3}}}{\left(D^2-4 D+3\right)^2}.
\end{equation}
In the same way one can expand the angular velocity $\Omega_c$ (given by (\ref{angularvel})) of the null geodesics in terms of $\alpha$:
$$ \Omega_c =  \sqrt{\frac{D-3}{D-1}} \left(\frac{D-2}{(D-1) \mu }\right)^{\frac{1}{D-3}} + $$
\begin{equation}
 \frac{2 \alpha  \left(\frac{D-2}{(D-1) \mu }\right)^{\frac{3}{D-3}}}{\sqrt{D-3} (D-1)^{3/2}} + {\cal O}(\alpha^2)
\end{equation}
Using (\ref{GenLyap}) we can show that the principal Lyapunov exponents has the  form
$$ \lambda =\frac{(D-3) \left(\frac{D-2}{(D-1) \mu }\right)^{\frac{1}{D-3}}}{\sqrt{D-1}}-\frac{2   \mu  \left(\frac{D-2}{(D-1) \mu
   }\right)^{\frac{D}{D-3}}}{\sqrt{D-1}} \alpha + $$
\begin{equation}\label{Lyapunov-explic}
\frac{2 (3 (D-8) D+28)  \left(\frac{D-2}{(D-1) \mu }\right)^{\frac{5}{D-3}}}{(D-3)
   (D-1)^{5/2}} \alpha ^2 +O\left(\alpha ^3\right).
\end{equation}
Here we expanded the Lyapunov exponents until the second order in $\alpha$, because there will be situation in which the difference between $\lambda$ and the eikonal quaisnormal modes appears only at the second order.
Notice, that the Lyapunov exponents are not invariant measures and should be interpreted with care \cite{Cornish:2003ig}.

\section{Gravitational perturbations of the Einstein-Lovelock black hole}

The natural generalization of the second order in curvature Gauss-Bonnet term to arbitrary order is given by the Loevlock theory \cite{Lovelock:1971yv}. A static spherically symmetric black-hole solution in the Einstein-Lovelock gravity is given by the general form (\ref{metric}), where (see \cite{Boulware:1985wk}, \cite{Wheeler:1985qd})
\begin{equation}\label{Lfdef}
f(r)=1-r^2\,\psi(r).
\end{equation}
The function $\psi(r)$ satisfies the following relation
$$ W(\psi(r))\equiv $$
\begin{equation}\label{LWdef}
\frac{D-2}{2}\left(\psi(r)+\sum_{m=2}^\infty\a_m\psi(r)^m\right)  = \frac{\mu}{r^{D-1}}\,,
\end{equation}
where
$$\a_m=\frac{\alpha_m}{m}\prod_{p=1}^{2m-2}(D-2-p)=\frac{\alpha_m}{m}\frac{(D-3)!}{(D-1-2m)!},$$
and $\a_m=0$ for any $D-2\leq2m$, implying that $W(\psi)$ is a finite polynomial of $\psi$.
Here we are interested only in the solutions to the above algebraic equations which describe the branch having the Einsteinian limit. In other words, we require that our black-hole metric goes over into the corresponding Tangherlini metric \cite{Tangherlini:1963bw} when $\alpha_m \rightarrow 0$.

Following \cite{Takahashi:2010ye}, we shall define a new function $T(r)$ as:
$$ T(r)\equiv r^{D-3}\frac{dW}{d\psi}= $$
\begin{equation}\label{LTdef}
\frac{(D-2) r^{D-3}}{2}\left(1+\sum_{m=2}^\infty m\a_m\psi(r)^{m-1}\right).
\end{equation}

The gravitational perturbation equations can be treated separately for irreducible representations, so that scalars, vectors and tensors relatively the $(D-2)$-dimensional rotation group obey separate sets of equations. In \cite{Takahashi:2010ye} it was shown that after the decoupling of the angular variables, the perturbations equations are reduced to the corresponding second-order master differential equations
\begin{equation}
\left(\frac{\partial^2}{\partial t^2}-\frac{\partial^2}{\partial r_*^2}+V_i(r_*)\right)\Psi_{i}(t,r_*)=0,
\end{equation}
where $\Psi_i$ are the wave functions for each type of perturbation: scalar, vector and tensor. In the eikonal regime, the effective potentials for all three types of gravitational perturbations can be approximated as follows
$$ V_t(r)=\ell^2\left(\frac{f(r)T''(r)}{(D-4)rT'(r)}+{\cal O}\left(\frac{1}{\ell}\right)\right),$$
\begin{equation} \label{dominant}
V_v(r)=\ell^2\left(\frac{f(r)T'(r)}{(D-3)rT(r)}+{\cal O}\left(\frac{1}{\ell}\right)\right),
\end{equation}
\begin{equation}\nonumber
V_s(r)=\ell^2\left(\frac{f(r)(2T'(r)^2-T(r)T''(r))}{(D-2) rT'(r)T(r)}+{\cal O}\left(\frac{1}{\ell}\right)\right).
\end{equation}
For further calculations it is useful to re-write the above formulas for the effective potentials symbolically as
\begin{equation}
V_{i} = \ell^2 \left(\frac{f_{i}(r)}{r^2} + {\cal O}\left(\frac{1}{\ell}\right)\right),
\end{equation}
where, $i$ stands for tensor (t), vector (v)  and scalar (s) types of gravitational perturbations. Thus,
$$ f_{t}(r) = \frac{f(r)r T''(r)}{(D-4)T'(r)}, \quad f_{v}(r) = \frac{f(r)r T'(r)}{(D-3)T(r)}, $$
\begin{equation}
f_{s}(r) = \frac{rf(r)(2T'(r)^2-T(r)T''(r))}{(D-2) T'(r)T(r)}.
\end{equation}
\begin{figure}
\resizebox{0.9 \linewidth}{!}{\includegraphics*{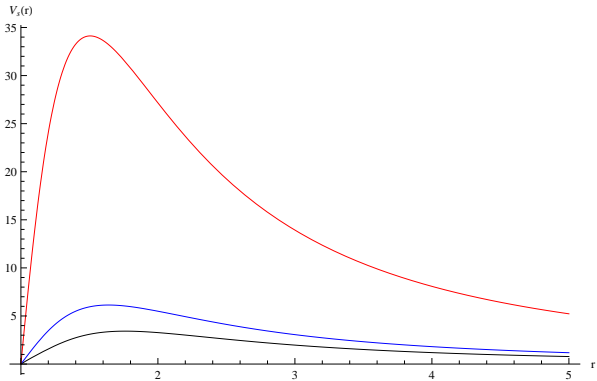}}
\caption{Effective potentials for the scalar-type gravitational perturbations of the asymptotically flat Einstein-Gauss-Bonnet black hole. Here, the black-hole radius $r_{H} =1$, $D=6$, $\alpha =1/10$ $\ell=3$ (black, bottom), $\ell=4$ (blue), $\ell=10$ (red, top).}\label{GBpotential}
\end{figure}
At high $\ell$, once the effective potential has the form of the potential barrier, falling off at the event horizon and spacial infinity, the WKB formula found in \cite{Schutz:1985zz} (for improvements and extensions of this formula, see \cite{Iyer:1986np,Konoplya:2003ii,Matyjasek:2017psv}) can be applied for finding quasinormal modes:
\begin{equation}
\frac{Q_0(r_0)}{\sqrt{2Q_0^{(2)}(r_0)}}=i(n+1/2). \label{wkb}
\end{equation}
Here, the second derivative $Q_0^{(2)}\equiv d^2Q_0/dr_*^2$ is evaluated at the extremum $r_{0}$ of the function $Q_0$. An example of such a ``good'' effective potential is shown on fig. (\ref{GBpotential}). Notice that in the Einstein-Lovelock theory such behavior of the potential barrier takes place only for sufficiently small values of the coupling constants, which correspond to the stable black hole. Otherwise, the effective potential may have a negative gap near the event horizon, which becomes deeper when $\ell$ is increased. It is important that in the eikonal regime $\ell \rightarrow \infty$ the WKB formula (\ref{wkb}) for potentials, like the one in fig. (\ref{GBpotential}), is \emph{exact}.
In the eikonal limit for each type of perturbations
\begin{equation}
Q_0\simeq \omega^2-f_{i}\frac{l^2}{r^2}
\end{equation}
Then we observe that
\begin{equation}\label{extremum}
2 f_{i}(r_0)=r_0 f_{i}'(r_0),
\end{equation}
i.e. as $f(r)$ does not coincide with $f_{i}(r)$, then the position of the effective potential's extremum $r_0$ must not coincide (in the general case) with the location of the null circular geodesic $r_c$. The WKB formula for quasinormal modes is also different from the Einsteinian ones, as now it includes $f_{i}(r)$ instead of $f(r)$:
\begin{equation}\label{main}
\omega_{\rm QNM i}=\ell \sqrt{\frac{f_{i0}}{r_0^2}}
-i\frac{(n+1/2)}{\sqrt{2}}
\sqrt{-\frac{r_0^2}{f_{i0}}\,\left (\frac{d^2}{dr_*^2}\frac{f_{i}}{r^2}\right )_{r_0}}.
\end{equation}
Thus, it is evident that, even when the effective potential has the form of the barrier, i. e. the WKB formula (\ref{wkb}) can be applied and is exact, in the general case:
\begin{itemize}
\item Radius of the circular null geodesics $r_{c}$ does not coincide with the position of the extremum of the effective potential $r_{i 0}$ in the eikonal regime;
\item The WKB formula for quasinormal modes include now the functions $f_{i}$ which are not identical to $f(r)$, so that eikonal quasinormal frequencies are different for each type of gravitational perturbations (scalar, vector, tensor) and different from the ones expected for the test scalar field.
\end{itemize}
Each of the above two reasons is sufficient for the breakdown of the proposed correspondence. Thus, it is evident from our general consideration of the Lovelock black holes that the characteristics of the null geodesics and eikonal quasinormal modes are not necessarily linked by the formula (\ref{QNM}). In the next section we shall write down analytical formulas for the eikonal quasinormal modes in terms of parameters of the Einstein-Gauss-Bonnet black holes and show the discrepancy between QN modes and null geodesics explicitly.

\section{Eikonal quasinormal modes in the Einstein-Gauss-Bonnet theory}

Here we shall derive analytical expressions for quasinormal modes in the regime of large multipole number $\ell$ for all three types of gravitational perturbations of the Einstein-Gauss-Bonnet black hole.

\textbf{Tensor type.}
Let us start from finding the position $r_0$ of the extremum of the effective potential, which can be expanded in terms of small $\alpha$:
\begin{equation}\label{rexpansion2}
r_{0} = r_{00} + r_{01} \alpha + r_{02} \alpha^2 +  {\cal O}(\alpha^3).
\end{equation}
Then, eq. (\ref{extremum}) expanded in $\alpha$ gives us the values of the coefficients $r_{0i}$. Thus, for the tensor type of perturbations one has
\begin{equation}\label{r00}
r_{00} = \left(\frac{D-2}{(D-1) \mu }\right)^{\frac{1}{3-D}},~ r_{01} =-\frac{4 (2 D-5) \left(\frac{D-2}{(D-1) \mu }\right)^{\frac{1}{D-3}}}{D^3-8 D^2+19 D-12},
\end{equation}
\begin{equation}\nonumber
r_{02} = \frac{8 (D (D (D (2 D-19)+49)+5)-64) \left(\frac{D-2}{(D-1) \mu }\right)^{\frac{3}{D-3}}}{(D-3)^2 \left(D^2-5 D+4\right)^2}.
\end{equation}
In a similar way one can find coefficients for the two other types of gravitational perturbations. From the above we can see that while $r_{00}$ given by (\ref{r00}) coincides with $r_{c0}$ given by (\ref{rc0}), that is not so for $r_{01}$ and $r_{c1}$ and all the higher corrections. In other words, while the positions of the null circular orbit and extremum of the effective potential coincide in the $D$-dimensional Schwarzschild space-time, they do not when the $\alpha$-correction is turned on.

Expanding the real part of (\ref{main}) in $\alpha$, one can see that the real oscillation frequency over the multipole number $\ell$ is
$$ \frac{Re (\omega)}{\ell} = \sqrt{\frac{D-3}{D-1}} \left(\frac{D-2}{(D-1) \mu }\right)^{\frac{1}{D-3}}+  $$
\begin{equation}
 \frac{6 (D-2) \sqrt{\frac{D-3}{D-1}}   \left(\frac{D-2}{(D-1) \mu
   }\right)^{\frac{3}{D-3}}}{D^3-8 D^2+19 D-12} \alpha +O\left(\alpha ^2\right),
\end{equation}
while the damping rate, characterised by  $Im (\omega)$, obeys the relation:

\begin{widetext}
$$  \frac{Im (\omega)}{ \left(n+\frac{1}{2}\right)}  = \frac{(D-3) \left(\frac{D-2}{(D-1) \mu }\right)^{\frac{1}{D-3}}}{\sqrt{D-1}}- \frac{2 \mu  \left(\frac{D-2}{(D-1) \mu
   }\right)^{\frac{D}{D-3}}}{\sqrt{D-1}} \alpha + $$
\begin{equation}
\frac{2 (D (D ((D-4) D (4 D-21)+144)-616)+484)  \left(\frac{D-2}{(D-1) \mu
   }\right)^{\frac{5}{D-3}}}{(D-4)^2 (D-3) (D-1)^{5/2}} \alpha ^2 +O\left(\alpha ^3\right)
\end{equation}
\end{widetext}
It is interesting to notice that while $Re (\omega)/\ell$ differs from the one expected from the angular velocity of null geodesics already in the linear order in $\alpha$, the value of $Im (\omega)/(n+1/2)$ differs from the  Lyapunov exponent, given by (\ref{Lyapunov-explic}),  only at the second  and higher orders in $\alpha$.

\textbf{Vector type.}
In a similar fashion, the position of the extremum of the effective potential is given by:
$$ r_{00} = \left(\frac{D-2}{(D-1) \mu }\right)^{\frac{1}{3-D}}, \quad r_{01} = \frac{2 \left(\frac{D-2}{(D-1) \mu }\right)^{\frac{1}{D-3}}}{D-1},$$
\begin{equation}
 r_{02} = -\frac{2 (D ((D-2) D-23)+36) \left(\frac{D-2}{(D-1) \mu }\right)^{\frac{3}{D-3}}}{\left(D^2-4 D+3\right)^2}.
\end{equation}
The real oscillation frequency $Re (\omega)$ obeys the following relation

$$ \frac{Re(\omega)}{\ell} = \sqrt{\frac{D-3}{D-1}} \left(\frac{D-2}{(D-1) \mu }\right)^{\frac{1}{D-3}}- $$
\begin{equation}
 \frac{2 \alpha  (D-2) \sqrt{\frac{D-3}{D-1}} \left(\frac{D-2}{(D-1) \mu
   }\right)^{\frac{3}{D-3}}}{D^2-4 D+3}+O\left(\alpha ^2\right).
\end{equation}
The relation for the damping rate $Im (\omega)$  reads
$$ \frac{Im (\omega)}{ \left(n+\frac{1}{2}\right)}  = \frac{(D-3) \left(\frac{D-2}{(D-1) \mu }\right)^{\frac{1}{D-3}}}{\sqrt{D-1}}-\frac{2 \alpha  \mu  \left(\frac{D-2}{(D-1) \mu
   }\right)^{\frac{D}{D-3}}}{\sqrt{D-1}}-  $$
\begin{equation}
\frac{2 \alpha ^2 (D ((D-18) D+51)-41) \left(\frac{D-2}{(D-1) \mu }\right)^{\frac{5}{D-3}}}{(D-3)
   (D-1)^{5/2}}+O\left(\alpha ^3\right).
\end{equation}

Here, again, we see that the damping rate differs from the one expected from the Lyapunov exponent only at the second and higher orders  in $\alpha$.

\textbf{Scalar type.}
The position of the extremum of the effective potential is given by:
\begin{equation}\nonumber
r_{00} = \left(\frac{D-2}{(D-1) \mu }\right)^{\frac{1}{3-D}}, \quad r_{01} = \frac{4 (D-2) \left(\frac{D-2}{(D-1) \mu }\right)^{\frac{1}{D-3}}}{D^2-4 D+3},
\end{equation}
\begin{equation}\nonumber
r_{02} =-\frac{8 (D (D ((D-6) D-5)+39)-32) \left(\frac{D-2}{(D-1) \mu }\right)^{\frac{3}{D-3}}}{(D-2) \left(D^2-4 D+3\right)^2}.
\end{equation}
The real oscillation frequency $Re (\omega)$ obey the relations:
$$ \frac{Re(\omega)}{\ell} = \sqrt{\frac{D-3}{D-1}} \left(\frac{D-2}{(D-1) \mu }\right)^{\frac{1}{D-3}}- $$
\begin{equation}
\frac{2 \alpha  \sqrt{\frac{D-3}{D-1}} (2 D-3) \left(\frac{D-2}{(D-1) \mu
   }\right)^{\frac{3}{D-3}}}{D^2-4 D+3}+O\left(\alpha ^2\right).
\end{equation}
The damping rate $Im (\omega)$, again can be found as a series expansion in small $\alpha$:
$$ \frac{Im (\omega)}{ \left(n+\frac{1}{2}\right)}  =  \frac{(D-3) \left(\frac{D-2}{(D-1) \mu }\right)^{\frac{1}{D-3}}}{\sqrt{D-1}}-\frac{2 \alpha  \mu  \left(\frac{D-2}{(D-1) \mu
   }\right)^{\frac{D}{D-3}}}{\sqrt{D-1}}+ $$
$$ \frac{2 (D (3 D (13 D-54)+232)-116) \alpha ^2 \left(\frac{D-2}{(D-1) \mu }\right)^{\frac{5}{D-3}-1}}{(D-3)
   (D-1)^{7/2} \mu } $$
\begin{equation}
+O\left(\alpha ^3\right).
   \end{equation}

The damping rate of the scalar type of gravitational perturbations differs from those of vector and tensor ones, again, only at the second and higher orders in $\alpha$. Notice that the higher order correction one wishes to find for the QN frequencies, the higher orders he needs to reach in the expansion of the position of the extremum of the effective potential.

\textbf{A test scalar field.}
For a test scalar field in the background of the Einstein-Gauss-Bonnet or Einstein-Lovelock black hole, the dominant centrifugal term in the effective potential is simply $f (r) \ell (\ell+1)/r^2$, so that up to a different function $f(r)$ (which now includes Lovelock coupling constants) all the deductions of \cite{Cardoso:2008bp} are strict and valid at all steps. Thus, the quasinormal frequencies of the test scalar field will evidently satisfy (\ref{QNM}), while the frequencies of gravitational perturbations are different for all three types and are different from those for the test scalar field even in the eikonal regime.

One should also remember that all the above formulas are obtained in the dominant (in terms of $1/\ell$-expansion) order of the eikonal regime. In order to use it for accurate estimations of quasinormal modes with sufficiently low $\ell$, one must take into consideration the next order of the $1/\ell$-expansion everywhere.  At $\alpha =0$, we reproduce the eikonal formulas found for the D-dimensional Schwarzschild black holes \cite{Konoplya:2003ii}. However, we used here different units and in order to reproduce, for example, eqs. (12, 13) of \cite{Konoplya:2003ii}, one should take in our formulas $\mu \rightarrow  (1/2)(D-2) \mu $.

\section{Discussion}

Though perturbations and quasinormal modes of black holes and branes in the Einstein-Gauss-Bonnet and Lovelock theories were considered in a number of papers \cite{LovelockQNM,Brigante:2007nu,Grozdanov} for various types of asymptotical behavior (flat, dS, AdS), no explicit analytical formula for the eikonal quasinormal frequencies of the gravitational perturbations of asymptotically flat black hole was presented. At the same time, the eikonal regime is special in Gauss-Bonnet and Lovelock theories, because, at sufficiently large values of coupling constants it bring a special kind of instability. Here we have found analytical expressions for quasinormal modes of gravitational perturbations of the Einstein-Gauss-Bonnet black hole.  When the Gauss-Bonnet coupling constants approaches zero, the analytic expressions for $\omega$ obtained here describe eikonal quasinormal modes of the D-dimensional Schwarzschild black hole. The gravitational quasinormal modes coincide with those for a test scalar field \cite{Konoplya:2003ii},\cite{Konoplya:2003dd} \emph{only} in the Einsteinian limit.

In our opinion, the broad belief that the eikonal quasinormal modes and unstable null geodesics are necessarily linked by eq. (\ref{QNM}) at least for any spherically symmetric stationary black holes, brought a number of misinterpretations in the current literature. For example, ``the universal upper bound'' for quasinormal modes of arbitrary spherical black holes  given by eq. 39 in \cite{Hod}, as well as its extension to the Gauss-Bonnet theory suggested by eq. 69 in \cite{Gallo}, does not take into account possibility of different features of the eikonal regime of gravitational perturbations, so that their arguments are compulsory only for test fields.

Therefore, a clear determination of the boarders of such an association between the two phenomena must have been spoken out.
Here we have learnt that although the association of the null geodesics with eikonal quasinormal modes exists in some cases, the range of its applicability is considerably constrained. Namely, the correspondence can be guaranteed for any stationary, spherically symmetric, asymptotically flat black holes only provided the two following conditions are fulfilled:

\begin{itemize}
\item Perturbations are described by a ``good'' (from the WKB point of view developed in \cite{Schutz:1985zz}) effective potential, i. e. the  potential barrier with a single extremum, implying the two turning points and decaying at the event horizon and infinity.
\item One is limited by perturbations of test fields only, and not of the gravitational field itself or other fields, which are non-minimally coupled to gravity.
\end{itemize}

In principle, the first condition must be satisfied for a test field in the background of a black-hole with well defined horizon, once $f(r)$ is positive everywhere outside the event horizon, so that the second condition alone is sufficient. This may be not true for more exotic objects, such as wormholes, naked singularities etc.

Rather unexpectedly, we have found that the damping rate of all three types of gravitational perturbations differs from that of a test scalar field (and consequently from the one predicted by the Lyapunov exponent) only at the second and higher orders of the Gauss-Bonnet coupling $\alpha$. This certainly cannot be interpreted on behalf of the correspondence, first, because it concerns only the imaginary part of $\omega$, and, second, because the relatively small difference for the $Im(\omega)$ at small $\alpha$ simply means that the damping rate is less sensitive to small curvature corrections than $Re(\omega)$.
It would be interesting to find analytical expressions for eikonal quasinormal modes in terms of black-hole parameters in the most general case of the Lovelock theory in a similar way it was done here for the Gauss-Bonnet black hole. However, as in the general case even the metric coefficients cannot be easily written in the explicit form, the final expressions may appear to be too much involved. Notice, that perturbations of a black hole in the nonlinear electrodynamics also show the non-standard behavior in the eikonal regime \cite{Chaverra:2016ttw}, so that it would be reasonable to check whether the correspondence works in this case. At the same time, for example, when analyzing test fields in the conformal gravity \cite{Toshmatov:2017bpx}, the correspondence is fulfilled according to our conclusions above.

\acknowledgments{R. K. was supported by ``Project for fostering collaboration in
science, research and education'' funded by the Moravian-Silesian Region, Czech Republic and by the Research Centre for Theoretical Physics and Astrophysics, Faculty of Philosophy and Science of Sileasian University at Opava. Z. S. acknowledges  the Albert Einstein Centre for Gravitation and Astrophysics supported under the Czech Science Foundation (Grant No. 14-37086G)}

\end{document}